\documentclass[pdflatex,sn-mathphys-num]{sn-jnl}

\usepackage{graphicx}
\usepackage{multirow}
\usepackage{amsmath,amssymb,amsfonts}
\usepackage{amsthm}
\usepackage{mathrsfs}
\usepackage[title]{appendix}
\usepackage{xcolor}
\usepackage{textcomp}
\usepackage{manyfoot}
\usepackage{booktabs}
\usepackage{algorithm}
\usepackage{algorithmicx}
\usepackage{algpseudocode}
\usepackage{listings}

\theoremstyle{thmstyleone}

\theoremstyle{thmstyletwo}

\theoremstyle{thmstylethree}

\begin{document}

\title[Article Title]{Phenomenological geometric ordering in fractional quantum Hall systems}

\author*{\fnm{M. A.} \sur{Hidalgo}}\email{miguel.hidalgo@uah.es}

\affil{\orgdiv{Departamento de Física y Matemáticas}, \orgname{Universidad de Alcalá}, \orgaddress{, \city{Alcalá de Henares}, \postcode{28805}, \state{Madrid}, \country{Spain}}}

\abstract{The fractional quantum Hall effect (FQHE) is conventionally understood in terms of strongly correlated many-body states and emergent quasiparticles with fractional charge. Here, we propose a complementary phenomenological framework in which impurity-induced geometric correlations within a Landau level contribute to the organization and stability of fractional quantum Hall states. The model considers a two-dimensional electron gas coupled to a correlated distribution of ionized impurities located at a finite distance from the electronic layer. Impurity-induced overlap between displaced Landau orbitals generates coherent guiding-center correlations and an effective splitting of the Landau-level degeneracy into fractional sublevels.

Within this framework, the resulting energy spectrum reproduces the principal odd-denominator fractional sequences through the interplay between guiding-center quantization and impurity-induced orbital coherence. An explicit expression for the correlation energy is obtained in terms of the magnetic length, impurity spacing, and impurity-layer separation, providing a direct connection between the proposed mechanism and experimentally controllable heterostructure parameters. The model naturally incorporates the integer quantum Hall regime as the limiting case of vanishing correlation-induced splitting.

We further discuss the relation of the proposed framework to several experimental signatures of the FQHE, including activated transport, shot-noise measurements, interferometric phase accumulation, and edge-state transport. Within this geometric picture, effective fractional factors emerge from collective orbital coherence and correlation-modified guiding-center dynamics rather than being uniquely associated with independent fractionally charged quasiparticles. Although the model does not attempt to derive topological order, anyonic statistics, or many-body incompressibility, it suggests that impurity-induced geometry and guiding-center coherence may provide an additional contribution to the emergence, stability, and experimental visibility of fractional quantum Hall states.}

\keywords{Two-dimensional electron systems, fractional quantum Hall effect}

\maketitle

\section{Introduction}

The quantum Hall effect, discovered experimentally by von Klitzing \cite{klitzing1980}, constitutes one of the clearest manifestations of quantum coherence in low-dimensional electronic systems. In a two-dimensional electron gas subjected to a strong perpendicular magnetic field, electronic motion becomes quantized into highly degenerate Landau levels, giving rise to the integer quantum Hall effect (IQHE), first observed experimentally by von Klitzing \cite{klitzing1980}, through Landau quantization and disorder-induced localization \cite{laughlin1981quantized}. The subsequent discovery of the fractional quantum Hall effect (FQHE) revealed incompressible electronic states at fractional filling factors \cite{laughlin1983anomalous}. The standard theoretical description is based on strongly correlated many-body states and emergent quasiparticles carrying fractional charge \cite{laughlin1983anomalous,jain1992microscopic}.

In the present work, we do not attempt to replace interaction-based descriptions of the fractional quantum Hall effect. Instead, we explore whether impurity-induced geometric correlations within a Landau level may provide an additional phenomenological contribution to the observed hierarchy of fractional states. In this picture, correlated impurity environments partially reorganize the effective Landau-level degeneracy through coherent orbital correlations associated with guiding-center geometry.

Despite the success of interaction-based approaches, several experimental features suggest that geometric aspects of Landau-level structure may also play an important role in fractional-state formation. In particular, the continuity between integer and fractional magnetotransport regimes and the strong dependence of fractional-state stability on sample quality, disorder, and heterostructure geometry indicate that spatial correlations within a Landau level may contribute significantly to the observed hierarchy of fractional states.

In addition, shot-noise measurements in fractional quantum Hall systems have traditionally been interpreted as evidence for fractionally charged quasiparticles through the relation
\begin{equation}
S_I=2q_{\mathrm{eff}}I_B
\label{eq:shot noise}
\end{equation}
where $S_I$ is the current-noise spectral density, $I_B$ is the backscattered current, and $q_{\mathrm{eff}}$ is interpreted as an effective fractional charge \cite{dePicciotto1997,saminadayar1997}. Experimentally, however, shot noise measures the scale of current fluctuations rather than directly identifying the microscopic nature of the transported entities. This leaves open the possibility that the observed effective fractional factors may emerge from collective orbital coherence and correlation-modified transport dynamics.

Similarly, activated longitudinal transport in fractional quantum Hall states is commonly interpreted in terms of quasiparticle excitation gaps \cite{boebinger1985,du1993experimental}. However, the experimentally observed activation energies may also contain contributions associated with impurity-induced splittings between correlation-modified orbital sublevels.

We therefore propose a phenomenological geometric scenario in which impurity-induced spatial correlations partially reorganize the effective Landau-level degeneracy. We consider a two-dimensional electron gas coupled to a correlated distribution of ionized impurities located at a finite distance from the electronic plane. The resulting impurity-induced orbital coherence generates effective fractional energy sublevels within a Landau level. The observed odd-denominator hierarchy is reproduced through the interplay between guiding-center quantization, \cite{cohen2019quantum1}, and correlated cyclotron motion, while the resulting spectral organization depends directly on impurity geometry and layer separation.

Edge transport in fractional quantum Hall systems is commonly described in terms of chiral edge excitations associated with topological order \cite{wen1990,chang2003}. However, part of the observed edge-state phenomenology may also reflect the geometric reorganization of correlated guiding-center orbitals near the boundaries of the sample, suggesting that the fractional hierarchy may emerge continuously from the underlying integer-quantized Landau structure through impurity-induced geometric correlations.

\section{Geometric correlations within Landau levels}

We consider a two-dimensional electron system subjected to a perpendicular magnetic field $\mathbf{B}=B \,\hat{\mathbf e}_{z}$. Within the effective-mass approximation and using the symmetric gauge, $\mathbf{A}=B (-y/2,x/2,0)$, the single-particle Hamiltonian is
\begin{equation}
H_0 = \frac{1}{2m^*} \left( \mathbf{p} + e\mathbf{A} \right)^2
\label{eq:H0}
\end{equation}
where $m^{*}$ is the effective electronic mass. The corresponding Landau eigenstates are
\begin{equation}
\psi_n^m(\mathbf{r},\theta)=C_{n,m}\,r^{|m|}e^{-r^2/4\ell_B^2}
L_n^{|m|}\left(\frac{r^2}{2l_B^2}\right)e^{im\theta}\label{eq:wave}
\end{equation}
with energies
\begin{equation}
E_n =\hbar\omega_c \left(n+\frac{1}{2}\right)
\end{equation}
where $\omega_c = eB/m^*$ and

\begin{equation}
l_B=\sqrt{\frac{\hbar}{eB}}
\label{eq:lB}
\end{equation}
is the magnetic length.

In the symmetric gauge, the electron coordinates can be decomposed into cyclotron orbit and guiding-center contributions. For a fixed Landau level, the guiding-center radius is associated with an effective geometric quantization scale of the form \cite{cohen2019quantum1}
\begin{equation}
\langle r_{GC}^2 \rangle = q l_B^2
\label{eq:larmor}
\end{equation}
where $q$ is an odd integer \cite{cohen2019quantum1}.

To investigate the effect of spatial correlations, we consider a correlated distribution of ionized impurities located at a finite distance $\Delta$ from the two-dimensional electron gas. The full Hamiltonian is
\begin{equation}
H = H_0 + U^{(i)}(\mathbf{r})
\label{eq:fullH}
\end{equation}
with impurity potential
\begin{equation}
U^{(i)}(\mathbf{r})=-e^2\sum_j\left[|\mathbf r-\mathbf R_j|^2+\Delta^2\right]^{-1/2}
\label{eq:impurity}
\end{equation}

In the high-field regime, where the magnetic length becomes comparable to the impurity-correlation scale, the continuous translational symmetry of the cyclotron motion is partially broken by the correlated impurity environment, giving rise to an effective correlated orbital structure. We therefore introduce a characteristic correlation lenght
\begin{equation}
|\boldsymbol{\xi}| = \eta d_i
\label{eq:xi}
\end{equation}
where $d_i$ is the dominant impurity spacing and $\eta$ is a dimensionless correlation parameter, (see below). The correlated orbital displacement is assumed to occur along an effective local correlation direction, $\boldsymbol{\xi} = \eta d_i \,\hat{\mathbf e}_{\xi}$ and the electronic states can be described by correlated combinations of displaced Landau orbitals
\begin{equation}
\Phi_{\xi}(k_q,\mathbf{r})=\frac{1}{\sqrt{N}}\sum_{\alpha}e^{ik_q\alpha\xi}\psi_n^m(\mathbf{r}-\alpha\boldsymbol{\xi})
\label{eq:Phi}
\end{equation}
which provide an effective phenomenological representation of impurity-induced orbital coherence within a correlation-modified Landau level, where $\alpha$ is an integer index. Since the guiding-center 
radius scales as 
\begin{equation}
\lambda_q \sim {\sqrt{q}\,l_B}
\label{eq:lambdaq}
\end{equation}
the characteristic correlation wave vector is
\begin{equation}
k_q=\frac{2\pi}{\sqrt{q}\,l_B}
\label{eq:kq}
\end{equation}
A schematic representation of the impurity-induced correlations considered in the present model is shown in Fig.~1.

The correlated orbital basis introduced in Eq.~(\ref{eq:Phi}) formally resembles a tight-binding representation constructed from displaced Landau orbitals. 

To make the origin of the correlated spectrum explicit, we evaluate the matrix elements of the full Hamiltonian using the displaced Landau orbitals defined in Eq.~(\ref{eq:wave}). Considering neighboring correlated orbitals separated by the displacement vector $\boldsymbol{\xi}$, the relevant matrix elements are
\begin{equation}
\langle\psi_n^m(\mathbf{r}+\alpha\boldsymbol{\xi})|H|\psi_n^m\left(
\mathbf{r}+(\alpha\pm1)\boldsymbol{\xi}\right)\rangle
\label{eq:matrix0}
\end{equation}

Defining the displaced orbitals as
\begin{equation}
\psi_\alpha(\mathbf{r})\equiv\psi_n^m(\mathbf{r}-\alpha\boldsymbol{\xi})
\label{eq:displaced}
\end{equation}
the corresponding matrix element becomes
\begin{equation}
M_{\alpha,\alpha\pm1}=\int d^2r
\psi_\alpha^*(\mathbf{r})H\psi_{\alpha\pm1}(\mathbf{r})
\label{eq:matrix1}
\end{equation}

Since the displaced Landau orbitals remain eigenstates of the unperturbed Hamiltonian $H_0$,
\begin{equation}
H_0\psi_{\alpha\pm1}=E_n\psi_{\alpha\pm1}
\label{eq:eigen}
\end{equation}
the matrix element separates naturally into a kinetic contribution and an impurity-induced contribution,
\begin{equation}
M_{\alpha,\alpha\pm1}=E_nS(\xi)+\langle\psi_\alpha|U^{(i)}|\psi_{\alpha\pm1}\rangle
\label{eq:matrix2}
\end{equation}

where the overlap between neighboring displaced orbitals is
\begin{equation}
S(\xi)=\langle\psi_\alpha|\psi_{\alpha\pm1}\rangle
\label{eq:overlap}
\end{equation}

For Landau states separated by a distance $\xi$, the overlap possesses the characteristic Gaussian form
\begin{equation}
S(\xi)=\exp\left(-\frac{\xi^2}{4l_B^2}\right)
\label{eq:gaussian}
\end{equation}

In the high-field regime, the overlap decreases rapidly with increasing separation. Consequently, only nearest-neighbor correlated orbitals contribute significantly to the effective coupling. In the basis represented in Eq.~(\ref{eq:Phi}), the impurity-induced overlap between neighboring guiding-center states generates an effective hopping amplitude between correlated cyclotron orbits. 

The impurity-induced contribution is therefore dominated by the matrix element
\begin{equation}
V_{\alpha,\alpha\pm1}=\langle\psi_\alpha|U^{(i)}|\psi_{\alpha\pm1}\rangle=\int d^2r\psi_\alpha^*(\mathbf{r})U^{(i)}(\mathbf{r})\psi_{\alpha\pm1}(\mathbf{r})
\label{eq:Vmatrix}
\end{equation}

Assuming a correlated impurity environment, the impurity potential may be expanded around the dominant impurity spacing $d_i$ as
\begin{equation}
U^{(i)}(\mathbf{r})\sim U_0+\delta U(\mathbf{r})
\label{eq:expandU}
\end{equation}
where the spatially uniform term $U_0$ only renormalizes the Landau energy $E_n$, while the correlated spatial fluctuations generate an effective orbital coupling,
\begin{equation}
V_{\alpha,\alpha\pm1}\equiv t(\xi)
\label{eq:tunnel}
\end{equation}

Because neighboring displaced orbitals form bonding and antibonding combinations, the resulting two-state spectrum becomes
\begin{equation}
E_\pm=E_n\pm t(\xi)
\label{eq:bonding}
\end{equation}

Introducing the effective impurity-induced correlation energy
\begin{equation}
\gamma \equiv 2t(\xi)
\label{eq:gammadef}
\end{equation}
one obtains
\begin{equation}
E_\pm=E_n\pm\frac{\gamma}{2}
\label{eq:Esplit}
\end{equation}

The dominant impurity-induced matrix elements may therefore be written as
\begin{equation}
\langle\psi_n^m(\mathbf{r}+\alpha\boldsymbol{\xi})|H|\psi_n^m\left(\mathbf{r}+\alpha\pm1)\boldsymbol{\xi}\right)\rangle=\pm\frac{\gamma}{2}
\label{eq:coupling}
\end{equation}
which corresponds to the effective nearest-neighbor coupling between correlated cyclotron states.

Using the Gaussian overlap between displaced Landau orbitals together with the effective Coulomb interaction generated by the impurity distribution, the coupling amplitude scales approximately as
\begin{equation}
t(\xi) \sim \frac{e^2}{\sqrt{d_i^2+\Delta^2}} \exp\left(
-\frac{\xi^2}{4l_B^2} \right)
\label{eq:tgauss}
\end{equation}
and, then, taking into account Eq.~(\ref{eq:gammadef}),
\begin{equation}
\gamma \sim \frac{2e^2}{\sqrt{d_i^2+\Delta^2}}
\exp\left(-\frac{\xi^2}{4l_B^2}\right)
\label{eq:gammaexplicit}
\end{equation}

Using the correlation length introduced in Eq.~(\ref{eq:xi}), the correlation energy may be written explicitly as
\begin{equation}
\gamma \sim \frac{2e^2}{\sqrt{d_i^2+\Delta^2}} \exp\!\left(- \frac{\eta^2 d_i^2}{4l_B^2} \right).
\label{eq:gammaBDelta}
\end{equation}
Using Eq.~(\ref{eq:lB}), the correlation energy can be expressed explicitly in terms of the magnetic field, establishing a direct dependence of the impurity-induced correlation energy on the magnetic field, the characteristic impurity spacing, and the separation between the impurity layer and the two-dimensional electron gas. The correlation energy therefore increases with magnetic field through the enhanced overlap of neighboring guiding-center orbitals, while larger values of $d_i$ or $\Delta$ suppress the effective coupling.

The Coulomb factor reflects the characteristic impurity potential scale, whereas the exponential term originates from the overlap between neighboring displaced Landau orbitals. Eq.~(\ref{eq:gammaexplicit}) and Eq.~(\ref{eq:gammaBDelta}) show explicitly that the stability of the correlation-induced fractional sublevels is controlled by experimentally accessible heterostructure parameters. Therefore, the strongest fractional states are expected when the impurity spacing is comparable to the magnetic length and the impurity layer remains sufficiently close to the electronic system, maximizing the effective correlation energy $\gamma$.

The expectation value of the Hamiltonian over the correlated states of Eq.~(\ref{eq:Phi}) is then
\begin{equation}
E(k_q)=\langle \Phi_\xi|H|\Phi_\xi\rangle
\label{eq:Ek}
\end{equation}

Substituting Eq.~(\ref{eq:Phi}) into Eq.~(\ref{eq:Ek}), one obtains
\begin{equation}
E(k_q)=\frac{1}{N}\sum_{\alpha,\beta}e^{ik_q(\alpha-\beta)\xi}\left\langle\psi_\beta\middle|H\middle|\psi_\alpha\right\rangle
\label{eq:double}
\end{equation}
where $\psi_\alpha(\mathbf{r})$ is given by  Eq.~\eqref{eq:displaced}

Retaining only nearest-neighbor couplings the resulting eigenvalue problem becomes mathematically analogous to a one-dimensional tight-binding chain, leading naturally through Eq.~(\ref{eq:coupling}) to the dominant contribution
\begin{equation}
E(k_q)=E_n\mp\frac{\gamma}{2}\left(e^{ik_q\xi}+e^{-ik_q\xi}\right)
\end{equation}
which yields
\begin{equation}
E(k_q)=E_n\mp\gamma\cos(k_q\xi)
\label{eq:band}
\end{equation}

This expression indicates that impurity-induced orbital coherence effectively modulates the Landau-level degeneracy through correlated cyclotron coupling. Expanding the cosine term to lowest nontrivial order,

\begin{equation}
\cos(k_q\xi) \sim 1-\frac{(k_q\xi)^2}{2}
\end{equation}
the energy spectrum becomes
\begin{equation}
E(k_q) \sim E_n \mp\gamma \pm \frac{\gamma}{2}(k_q\xi)^2
\label{eq:expansion}
\end{equation}

Defining the shifted Landau energy
\begin{equation}
\tilde{E}_n=E_n\mp\gamma
\label{eq:shifted}
\end{equation}
and using Eq.~(\ref{eq:kq}), the impurity-induced correction becomes
\begin{equation}
\Delta E_q^\eta=\frac{\gamma}{2}\left(\frac{2\pi\eta d_i}{\sqrt{q} l_B} \right)^2
\label{eq:deltaE}
\end{equation}

To parametrize different families of correlated orbital configurations, we introduce the relation
\begin{equation}
\eta^2=p\
\label{eq:p}
\end{equation}
where the integer $p$ labels the dominant guiding-center correlation mode associated with a given fractional sequence, (see below).

Using Eqs.~(\ref{eq:deltaE}) and (\ref{eq:kq}), the impurity-induced correction scales as

\begin{equation}
\Delta E_q^\eta = \frac{\eta^2}{q}\hbar\omega_c=\frac{p}{q}\hbar\omega_c
\label{eq:DeltaE_q}
\end{equation}

Therefore
\[E_n+\Delta E_q^\eta=\left(n+\frac12\pm\frac{p}{2q}\right)\hbar\omega_c\]
which can be written as 

\begin{equation}
E=\left(2n+1\pm\frac{p}{q}\right)E_0
\label{eq:fractionalE}
\end{equation}
with
\begin{equation}
E_0=\frac{\hbar\omega_c}{2}
\label{eq:E0}
\end{equation}

An important consistency check of the model is that it naturally incorporates the integer quantum Hall regime, originally discovered by von Klitzing \cite{klitzing1980}. In the absence of impurity-induced orbital correlations, for vanishing correlation parameter, $p=0$, Eq.~(\ref{eq:fractionalE}) reduces to the ordinary Landau-level spectrum $E=(2n+1)E_0=\hbar\omega_c\left(n+\frac12\right)$, which corresponds to the conventional integer quantum Hall effect  \cite{hidalgo2021quantum}. 
The fractional hierarchy therefore appear naturally as a correlation-induced geometric reorganization of an underlying integer-quantized Landau spectrum. The model consequently provides a unified phenomenological framework connecting the integer and fractional quantum Hall regimes through impurity-induced guiding-center coherence.
In this sense, the impurity-induced correlations do not replace the Landau quantization responsible for the IQHE, but instead partially reorganize its degeneracy into correlation-modified fractional sublevels.

Eq.~(\ref{eq:fractionalE}) summarizes the effective fractional energy structure obtained within the model. The fractional hierarchy emerges from the interplay between guiding-center quantization and impurity-induced orbital correlations. Using our notation, the dominant experimentally observed odd-denominator states correspond to the shortest correlation lengths, namely $\eta=1$ and $\eta=\sqrt{3}$.

Moreover, the absence of a robust incompressible state at filling factor $1/2$ follows from the suppression of the geometric correlations responsible for the odd-denominator hierarchy, corresponding to the limit $p\rightarrow0$.

\subsection{Phenomenological assumptions of the model}

The model is intended as a phenomenological description of correlation-induced geometric reorganization within a Landau level. Several simplifying assumptions are introduced in order to isolate the possible role of impurity-induced guiding-center coherence in the emergence of fractional spectral structures.

First, the impurity environment is assumed to possess a partially correlated spatial distribution characterized by a dominant impurity spacing $d_i$ and a finite layer separation $\Delta$. These correlations generate effective overlap between displaced cyclotron orbitals.

Second, the correlated electronic states are constructed from coherent superpositions of displaced Landau orbitals, Eq.~(\ref{eq:Phi}), in a form analogous to a tight-binding representation. Because Landau orbitals exhibit Gaussian localization on the scale of the magnetic length, overlap contributions between distant orbitals become exponentially suppressed. Consequently, only nearest-neighbor correlated couplings are retained within the present approximation.

Third, the impurity-induced coupling parameter $\gamma$ is treated phenomenologically as an effective correlation energy controlled by the overlap between neighboring guiding-center states and by the geometry of the impurity distribution.

Finally, the model does not attempt to derive the microscopic many-body topological order conventionally associated with fractional quantum Hall states. Instead, the it explores whether impurity-induced orbital coherence and guiding-center geometry may contribute to the effective organization, stability, and visibility of experimentally observed fractional spectral structures.

Although the present derivation has been formulated in the symmetric gauge, the underlying physical mechanism is associated with guiding-center correlations and impurity-induced orbital coherence rather than with a specific gauge choice. The phenomenological fractional sublevel structure therefore reflects the correlated geometry of the Landau-level orbitals and is expected to remain gauge independent.

\section{Connection with experimental fractional sequences}

The impurity-induced splitting of the Landau spectrum modifies the density of states and generates additional minima in the longitudinal resistivity together with Hall plateaux at fractional filling factors. The spectrum is obtained through the substitution
\begin{equation}
E\rightarrow\left(2n+1\pm\frac{p}{q}\right)E_0
\label{eq:fractionalDOS}
\end{equation}

Experimentally, Shubnikov–de Haas minima occur when the Fermi level crosses the fractional sublevels, leading to the magnetic-field sequence
\begin{equation}
B_{n,p,q}\sim\frac{2m^*E_F}{\hbar e\left(2n+1\pm\frac{p}{q}\right)}
\label{eq:Bpq}
\end{equation}
where
\begin{equation}
E_F=\frac{2\pi\hbar^2 n_e}{m^*}
\end{equation}
is the Fermi energy of the two-dimensional electron gas with electron density $n_e$. 

In fact, using realistic GaAs/AlGaAs densities,
$n_e \simeq 3 \times 10^{15}\,\mathrm{m}^{-2}$, the resulting spectrum obtained suggests a possible geometric organization of the dominant odd-denominator fractions, in particular the experimentally robust $1/3$, $2/3$, $2/5$, and $3/5$ states
\cite{willett1987observation,clark1986odd,willett1988termination,du1993experimental,choi2008activation,shabani2010fractional}.

The corresponding impurity-induced sublevel structure associated with the $1/3$ hierarchy is illustrated schematically in Fig.~2.
Higher-order odd-denominator states emerge from longer guiding-center correlation scales and smaller effective sublevel splittings. The corresponding phenomenological organization associated with the $1/5$ family is shown in Fig.~3.

Within our framework, transport through a quantum point contact occurs through coherent orbital structures rather than through independent localized quasiparticles. The experimentally observed shot-noise relation \cite{bid2009,heiblum2003},  Eq.~(\ref{eq:shot noise}), may also admit a complementary geometric interpretation in terms of correlation-modified orbital coherence.

Since the correlated orbital states occupy an effective coherent area
\begin{equation}
A_q\sim \lambda_q^{2} \sim ql_B^2
\label{eq:Aq}
\end{equation}
transport through a quantum point contact is no longer carried by independent localized orbitals but by extended guiding-center coherent structures. For a fixed active transport area, A, the number of statistically independent fluctuating regions is therefore reduced to
$N_q \sim \frac{A}{A_q}=\frac{A}{q l_B^2}$. Compared with the uncorrelated case $(q=1)$, the number of independent fluctuation channels is reduced by the factor $\frac{N_q}{N_1}\sim\frac{1}{q}$
The shot-noise suppression factor is therefore naturally identified as
$F_q=\frac{N_q}{N_1}\sim\frac{1}{q}$. 
Equivalently, using the guiding-center coherence length,

\begin{equation}
F_q\sim\frac{l_B^2}{\lambda_q^2}\sim\frac1q
\label{eq:Cq}
\end{equation}
The reduction of the effective number of statistically independent transport channels originates from the spatial coherence of the correlated guiding-center structures rather than from the existence of independent fractionally charged carriers. Consequently, the experimentally measured Fano factor reflects the collective geometry of the correlated orbital states.
Then, the resulting noise spectrum becomes
\begin{equation}
S_I=2eI_B\,F_q\sim2eI_B\frac1q
\label{eq:noise2}
\end{equation}
The present argument should be regarded as a phenomenological scaling 
relation for fluctuation suppression and not as a microscopic derivation of the experimentally measured Fano factor.

Experimentally, Eq.~(\ref{eq:noise2}) is formally equivalent to the conventional expression obtained using an effective fractional charge $e/q$. Within the present picture, however, the observed Fano factor emerges from the spatial coherence scale of the correlated cyclotron states rather than from the existence of fractionally charged quasiparticles. The effective fractional factor therefore reflects the geometry of the correlated orbital structures and the reduced number of independent fluctuation channels associated with guiding-center quantization.

A characteristic phenomenological energy scale associated with the fractional sublevels is approximately Eq.~(\ref{eq:DeltaE_q}). This energy scale corresponds to the separation between impurity-induced correlated orbital sublevels inside a broadened Landau band. Thermally activated longitudinal transport therefore arises from transitions between these correlation-modified orbital states. The longitudinal conductivity is then expected to follow an activated behavior of the form
\begin{equation}
\sigma_{xx}\sim\exp\left(-\frac{\Delta E_{p/q}}{2k_B T}\right)
\label{eq:activated}
\end{equation}
consistent with the experimentally observed activated transport in fractional quantum Hall states.

Because the fractional gaps scale approximately as Eq.~(\ref{eq:DeltaE_q}) low-order odd-denominator fractions possess the largest activation energies and therefore the greatest experimental stability. Higher-order fractions exhibit smaller sublevel separations and become increasingly sensitive to disorder broadening and thermal fluctuations.

The activation gaps are further reduced by disorder broadening $\Gamma$, so that the experimentally observed effective gap may be written phenomenologically as
\begin{equation}
\Delta_{\mathrm{eff}}\simeq\Delta E_{p/q}-\Gamma
\label{eq:effectivegap}
\end{equation}

The strong experimental dependence of fractional quantum Hall states on sample mobility and disorder emerges naturally from the impurity-induced coherence mechanism itself. Since the correlated orbital states are formed through coherent coupling between displaced Landau orbitals, disorder broadening progressively suppresses the spatial coherence required for the formation of fractional sublevels.

As the disorder scale $\Gamma$ approaches the correlation-induced splitting $\Delta E_{p/q}$, the overlap between neighboring correlated orbital states becomes increasingly incoherent, leading to the progressive destruction of the fractional spectral organization. Consequently, only high-mobility samples with sufficiently small disorder broadening satisfy the stability condition 
\begin{equation}
\Gamma<\Delta E_{p/q}   
\label{eq:criterion1}
\end{equation}
consistent with the experimentally observed fragility of higher-order fractional states. Hence, the condition for experimental observability of a fractional state therefore becomes
\begin{equation}
\frac{p}{q} \gtrsim \frac{\Gamma}{\hbar\omega_c}
\label{eq:criterion}
\end{equation}
where $\Gamma$ denotes the disorder broadening. Eq.~(\ref{eq:criterion}) naturally explains the predominance of low-order odd-denominator fractions and the suppression of higher-order states as disorder increases.

On the other hand, the emergence of Hall plateaux at fractional filling factors is associated with the stabilization of correlation-induced sublevels inside the broadened Landau spectrum. When the Fermi level lies inside the effective gap between neighboring correlated sublevels, the longitudinal conductivity becomes strongly suppressed, while the transverse conductivity remains approximately quantized.

If the correlation-induced sublevels become sufficiently robust against disorder broadening, the resulting spectral organization may phenomenologically support Hall plateaux at effective filling factors
$\nu=\frac pq$ In that regime, it may phenomenologically reproduce the experimentally observed form, \cite{hidalgo2021quantum},
\begin{equation}
\sigma_{xy}=\nu\frac{e^2}{h}
\label{eq:sigmaxy}
\end{equation}

Then, the stability of the fractional Hall plateaux is controlled phenomenologically by the coherence of the correlated guiding-center structures and by the impurity-induced separation between fractional sublevels. In particular, the activation condition Eq.~(\ref{eq:criterion1}) ensures suppression of dissipative bulk transport and stabilization of the corresponding Hall plateau.

The model does not attempt to derive the exact topological quantization conventionally associated with fractional Hall conductance. Rather, it suggests that impurity-induced geometric correlations and guiding-center coherence may contribute to the effective organization and robustness of the experimentally observed fractional conductivity sequences.

The correlated orbital structures proposed may also influence quantum-interference phenomena observed in fractional quantum Hall interferometers \cite{camino2005,willett2009,nakamura2020}. Since the coherent states in Eq.~(\ref{eq:Phi}) extend over guiding-center correlation lengths given by Eq.~(\ref{eq:lambdaq}), electronic transport through interferometric geometries can accumulate collective orbital phases associated with correlation-modified cyclotron motion.
The phase accumulated by a charged particle moving around a closed trajectory enclosing a magnetic flux $\Phi$ is given by the Aharonov--Bohm relation
\begin{equation}
\phi=\frac{e}{\hbar}\oint \mathbf A\cdot d\mathbf l=2\pi\frac{\Phi}{\Phi_0}
\label{eq:AB}
\end{equation}

where
\begin{equation}
\Phi_0=\frac{h}{e}
\end{equation}

is the magnetic-flux quantum. The relevant enclosed area is associated with the coherent guiding-center structure. Since the characteristic correlated area scales as Eq.~(\ref{eq:Aq}), the enclosed magnetic flux becomes
\begin{equation}
\Phi_q = B A_q
\end{equation}

Substituting into Eq.~(\ref{eq:AB}) yields
\begin{equation}
\phi_q=2\pi\frac{B A_q}{\Phi_0}
\label{eq:phase}
\end{equation}
Using Eq.~(\ref{eq:Aq}) together with
\begin{equation}
l_B^2=\frac{\hbar}{eB}=\frac{\Phi_0}{2\pi B}
\end{equation}
one obtains
\begin{equation}
\phi_q=2\pi\frac{B\,q\,l_B^2}{\Phi_0} \sim q
\label{eq:phaseq}
\end{equation}

The interference periodicities therefore become directly linked to the geometrically correlated guiding-center area. Within this picture, part of the experimentally observed interference phenomenology may emerge from collective orbital coherence and correlation-modified guiding-center geometry. This phase relation should be regarded as a phenomenological geometric scaling argument rather than as a microscopic derivation of anyonic braiding statistics.

The present framework may also provide a complementary geometric interpretation of edge transport in fractional quantum Hall systems. Near the boundaries of the sample, the confining potential breaks the translational symmetry of the correlated orbital structures, leading to preferential guiding-center drift along the sample edge.

As a consequence, the coherent states defined in Eq.~(\ref{eq:Phi}) become spatially distorted near the boundary and generate quasi-one-dimensional correlated transport channels. The characteristic spatial extent of these channels is governed by the guiding-center coherence length [Eq.~(\ref{eq:lambdaq})], which determines the size of the correlated cyclotron motion.

A quantitative estimate follows directly from the geometric structure of the model. Since the correlated guiding-center orbitals extend over a distance of order $\lambda_q$, the effective width of the boundary transport channels is expected to scale as
\begin{equation}
w_q \sim \lambda_q \sim \sqrt{q} l_B
\label{eq:edgewidth}
\end{equation}

For a sample edge of length (L), the maximum number of independent coherent channels that can be accommodated along the boundary is therefore approximately
\begin{equation}
N_{\mathrm{edge}}
\sim
\frac{L}{w_q}=\frac{L}{\sqrt{q} l_B}
\label{eq:Nedge}
\end{equation}

These relations establish a direct connection between edge transport and the same guiding-center coherence scale that governs the impurity-induced fractional sublevel structure in the bulk. As the correlation parameter (q) increases, the coherent orbital structures occupy larger spatial regions, reducing the number of independent edge channels available along a given boundary.

Consequently, higher-order fractional states are expected to exhibit enhanced sensitivity to disorder and boundary perturbations, consistent with the experimentally observed fragility of weak fractional states. Within the present picture, the robustness of edge transport against moderate disorder follows from the collective coherence of the correlated orbital structures rather than exclusively from topological protection associated with fractionally charged quasiparticles.

The resulting chiral propagation reflects the combined effects of magnetic confinement, guiding-center quantization, and impurity-induced orbital coherence. In this sense, boundary-induced reorganization of correlated guiding-center orbitals may contribute to certain aspects of the experimentally observed edge-state phenomenology.

In conventional descriptions, fractional edge transport is formulated in terms of chiral Luttinger liquids \cite{wen1990}. The present model does not attempt to reproduce the full chiral Luttinger-liquid formalism or derive its associated critical exponents. Instead, it suggests that part of the observed edge phenomenology may be influenced by the geometric organization of impurity-correlated guiding-center orbitals, with the coherence length $\lambda_q$ providing the fundamental scale that controls both bulk fractional sublevel formation and edge-state transport.

\section{Order of magnitude estimates}

To assess the physical plausibility of the proposed impurity-induced fractional sublevel structure, we consider experimentally realistic GaAs/AlGaAs heterostructure parameters. For magnetic fields in the range $B\sim8-15 \mathrm{T}$, the magnetic length is $l_B\sim6-9\,\mathrm{nm}$. Typical modulation-doped heterostructures exhibit impurity-layer separations and characteristic impurity spacings in the range $d_i$, $\Delta\sim20-50 \mathrm{nm}$.
The dependence of the correlation energy on magnetic field, impurity spacing and impurity-layer separation provides a direct connection between the proposed mechanism and experimentally controllable heterostructure parameters. In particular, the model predicts that the impurity-induced coupling increases with magnetic field through the reduction of the magnetic length, while it is suppressed as either the impurity spacing $d_i$ or the impurity-layer separation $\Delta$ increase. The impurity-induced fractional energy scale is controlled by Eq.~(\ref{eq:deltaE}).

For low-order odd-denominator states and moderate correlation parameters $\eta=1$, the resulting energy corrections correspond to sub-meV to meV scales, comparable to experimentally observed disorder broadening and activation gaps in high-mobility samples \cite{boebinger1985,du1993experimental}.

These estimates suggest that impurity-induced orbital correlations may contribute appreciably to the effective organization, activated transport behavior, and visibility of fractional quantum Hall states under experimentally accessible conditions.

\section{Relation to interaction-based descriptions}

The standard theoretical description of the fractional quantum Hall effect is based on strongly correlated many-body states, including Laughlin wavefunctions and composite-fermion approaches, which successfully account for incompressibility, fractional charge, and topological properties of fractional Hall states \cite{laughlin1983anomalous,jain1992microscopic}. The present framework does not exclude the simultaneous presence of interaction-driven many-body correlations and fractionally charged quasiparticles. Instead, it explores whether impurity-induced geometric coherence may provide an additional organizational mechanism contributing to the effective fractional spectral structure observed experimentally. Then, the correlated impurity environments partially reorganize the effective orbital structure through guiding-center coherence and correlated cyclotron motion.

This viewpoint is also qualitatively consistent with previous works emphasizing the importance of guiding-center geometry and internal geometric degrees of freedom in quantum Hall systems \cite{haldane2011geometrical}. 

The broader idea that geometric organization and dimensional constraints may strongly influence quantum Hall phenomenology is also qualitatively consistent with previous studies of unconventional Hall responses in higher-dimensional correlated systems, including recent work on three-dimensional quantum Hall behavior in anisotropic electronic structures \cite{hidalgo2025approach}.

In the same spirit, it suggests that shot-noise measurements and activated transport gaps in the fractional quantum Hall regime may admit an alternative geometric interpretation. In this framework, the experimentally observed effective Fano factors and activation energies arise from collective orbital coherence, impurity-induced sublevel splittings, and correlation-modified transport channels rather than from the dynamics of independent fractionally charged quasiparticles.

The present framework should be viewed as a phenomenological orbital-geometry model for the partial reorganization of Landau-level spectral structure under correlated impurity environments. It is not intended to reproduce the full topological many-body structure of the fractional quantum Hall effect, including anyonic braiding statistics, topological degeneracy, or conformal edge theories. Rather, the model explores whether part of the experimentally observed fractional hierarchy and transport phenomenology may also contain an important geometric contribution associated with guiding-center coherence and impurity-induced orbital correlations.

\section{Discussion}

The present results suggest that impurity-induced geometric correlations can partially reorganize the Landau-level degeneracy and contribute to the emergence of fractional quantum Hall states. The stability of fractional states is directly linked to impurity correlations and heterostructure geometry, particularly to the separation $\Delta$ between the impurity layer and the two-dimensional electron gas.

The proposed mechanism is phenomenological and does not explicitly incorporate electron--electron interactions. The model is intended to identify possible geometric and disorder-induced contributions to the effective organization of fractional spectral structures within a Landau level, and should be regarded as complementary to interaction-based descriptions such as Laughlin and composite-fermion theories, The geometric interpretation of shot noise proposed here does not attempt to exclude conventional quasiparticle descriptions, but instead explores whether the experimentally observed effective fractional noise factors may emerge from collective orbital coherence associated with correlation-modified Landau states.

And the same phenomenological viewpoint also applies to the activated energy gaps observed experimentally in fractional quantum Hall transport. These gaps arise from impurity-induced splittings between correlated orbital sublevels and from the thermal activation required to overcome disorder-broadened coherence gaps inside a partially reorganized Landau level. With our picture, the activation energies therefore reflect the stability of the correlated orbital structures rather than isolated quasiparticle excitation energies.

The present framework may also provide an alternative geometric perspective on interferometric experiments in the fractional quantum Hall regime. In Fabry--Pérot and Mach--Zehnder geometries, the observed oscillatory transport patterns are commonly interpreted in terms of phase accumulation by fractionally charged anyonic quasiparticles \cite{camino2005,willett2009,nakamura2020}. However, similar interference effects may arise from the collective phase coherence of correlation-modified cyclotron states. Because the correlated orbital structures extend over guiding-center coherence lengths of order Eq.~\eqref{eq:lambdaq} the accumulated phase in an interferometric trajectory depends on the geometric organization of the correlated Landau orbitals and on the impurity-induced coherence pattern inside the device. The resulting interference periodicities may therefore reflect collective orbital coherence and guiding-center quantization without explicitly addressing the microscopic quasiparticle dynamics conventionally associated with anyonic interference. However, the model does not attempt to derive anyonic exchange statistics or topological braiding properties. Rather, it suggests that part of the experimentally observed interferometric phenomenology may be influenced by correlation-induced geometric coherence within partially reorganized Landau levels. This feature naturally explains the strong experimental sensitivity of fractional quantum Hall states to sample mobility, impurity concentration, heterostructure geometry, and disorder correlations. In particular, the rapid suppression of higher-order fractional states in lower-mobility samples follows directly from the reduction of coherence between correlated cyclotron orbits.

Concerning on the edge-state phenomenology \cite{wen1990,chang2003} some aspects may be influenced by collective guiding-center coherence and boundary-induced orbital reorganization. The model explores whether part of the experimentally observed edge transport may emerge from correlation-modified Landau-level geometry.

Moreover, an intriguing possibility is that the geometric phases generated by correlated guiding-center structures may constitute an effective mesoscopic manifestation of the same underlying physics that appears as fractional statistics in interaction-based descriptions. From this perspective, anyonic phases and impurity-induced guiding-center coherence would not necessarily represent mutually exclusive mechanisms, but rather complementary descriptions of different aspects of the correlated Landau-level geometry.

A possible experimental test would consist in systematically modifying the impurity correlation length and impurity-layer separation in otherwise comparable heterostructures. The effective fractional sublevel splitting is directly controlled by the impurity-induced correlation energy $\gamma$, implying that the stability and visibility of fractional states should exhibit a measurable dependence on impurity geometry beyond conventional mobility effects alone.

The model should be interpreted as a complementary geometric contribution to fractional-state organization rather than as a complete microscopic theory of the fractional quantum Hall effect.

\section{Conclusion}

The present geometric interpretation is qualitatively consistent with previous studies emphasizing the role of orbital geometry and dimensional effects in quantum Hall systems, including recent extensions of Hall quantization phenomenology to three-dimensional electronic structures \cite{hidalgo2025approach}.

We have presented a phenomenological geometric framework that may contribute to the organization of fractional quantum Hall states based on impurity-induced correlations within a Landau level. A correlated impurity environment generates coherent coupling between cyclotron orbits and provides fractional sublevels associated with odd-denominator filling factors.

The estimated energy scales and magnetic-field dependence are consistent with experimentally accessible regimes and predict a direct dependence of fractional-state stability on impurity geometry, disorder correlations, and layer separation. On the other hand, even-denominator states are not naturally stabilized.

Shot-noise measurements may also reflect collective orbital coherence and correlation-modified transport fluctuations associated with impurity-correlated guiding-center structures rather than as direct evidence for independent fractionally charged quasiparticles. Likewise, the experimentally observed activated transport gaps may be interpreted as arising from impurity-induced orbital sublevel splittings and thermally activated transitions between correlated Landau states.

The framework furthermore suggests that certain interferometric phenomena observed in fractional quantum Hall devices may admit a complementary interpretation in terms of collective orbital coherence and guiding-center geometry.

Moreover, the remarkable robustness of fractional Hall plateaux may partly reflect the collective stability of the impurity-correlated guiding-center structures under disorder and magnetic confinement.

Finally, the model connects the integer and fractional quantum Hall regimes through correlation-induced reorganization of Landau-level degeneracy.
Additionally, it suggests that controlled impurity engineering and heterostructure design may provide additional routes for tuning effective fractional spectral organization in high-mobility quantum Hall systems. Moreover, it also naturally explains the pronounced dependence of fractional quantum Hall states on sample mobility and disorder. Since the correlated fractional sublevels originate from impurity-induced orbital coherence, disorder broadening directly controls the stability and visibility of the resulting fractional structures.

\section{Data Availability}
The datasets generated and/or analysed during the current study, corresponding to the data used to produce Figures 2 and 3, are available in the supplementary information files associated with this published article.

\section{Funding}

The authors received \textbf{no funding} for this work.

\section{Figures}

\begin{figure}[ht]
\centering
  \includegraphics[width=0.75\columnwidth]{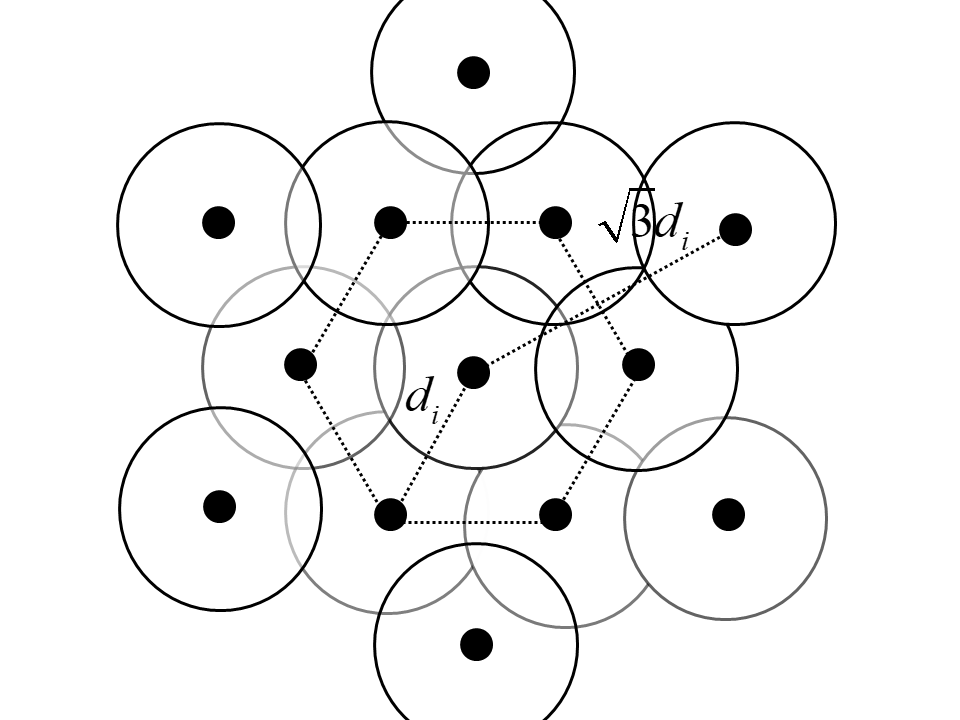}
  \caption{Schematic representation of impurity-induced correlations between displaced Landau orbitals. Zenithal view of the impurity plane and the two-dimensional electron system (2DES). The correlated guiding-center basic displacement, $(\boldsymbol{\xi})$, couples neighboring cyclotron states separated by the characteristic impurity spacing $(d_i)$, generating coherent orbital structures within a Landau level. The impurity distribution is highlighted in black.}
  \label{fig:figure1}
\end{figure}

\begin{figure}[ht]
\includegraphics[width=0.75\columnwidth]{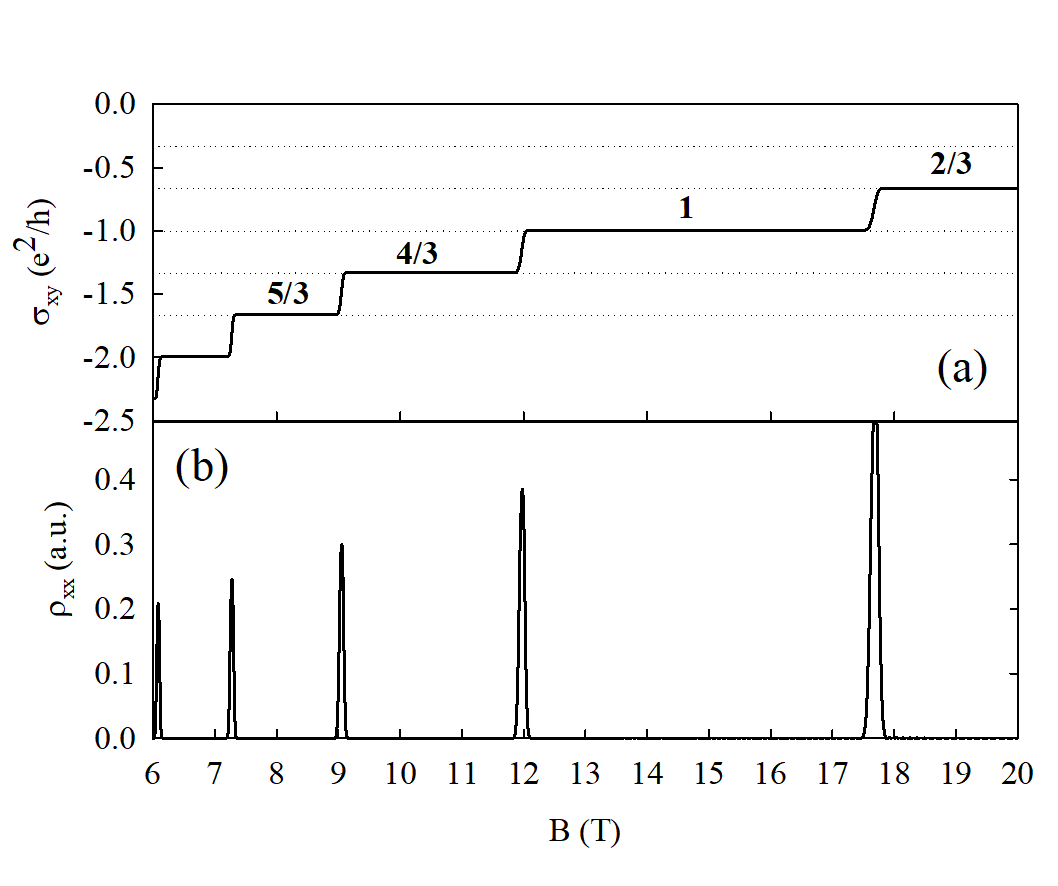}
\caption{Phenomenological impurity-induced fractional sublevel structure associated with the dominant $1/3$ hierarchy.
(a) Effective Hall magnetoconductivity exhibiting fractional plateau-like structures associated with the impurity-induced sublevels. (b) Corresponding longitudinal magnetoresistivity minima obtained from the fractional energy spectrum of Eq.~(\ref{eq:fractionalE}). The results illustrate how correlated orbital coherence can reorganize the effective Landau-level structure into experimentally relevant odd-denominator sequences.}
\label{fig:figure2}
\end{figure}

\begin{figure}[ht]
\includegraphics[width=0.75\columnwidth]{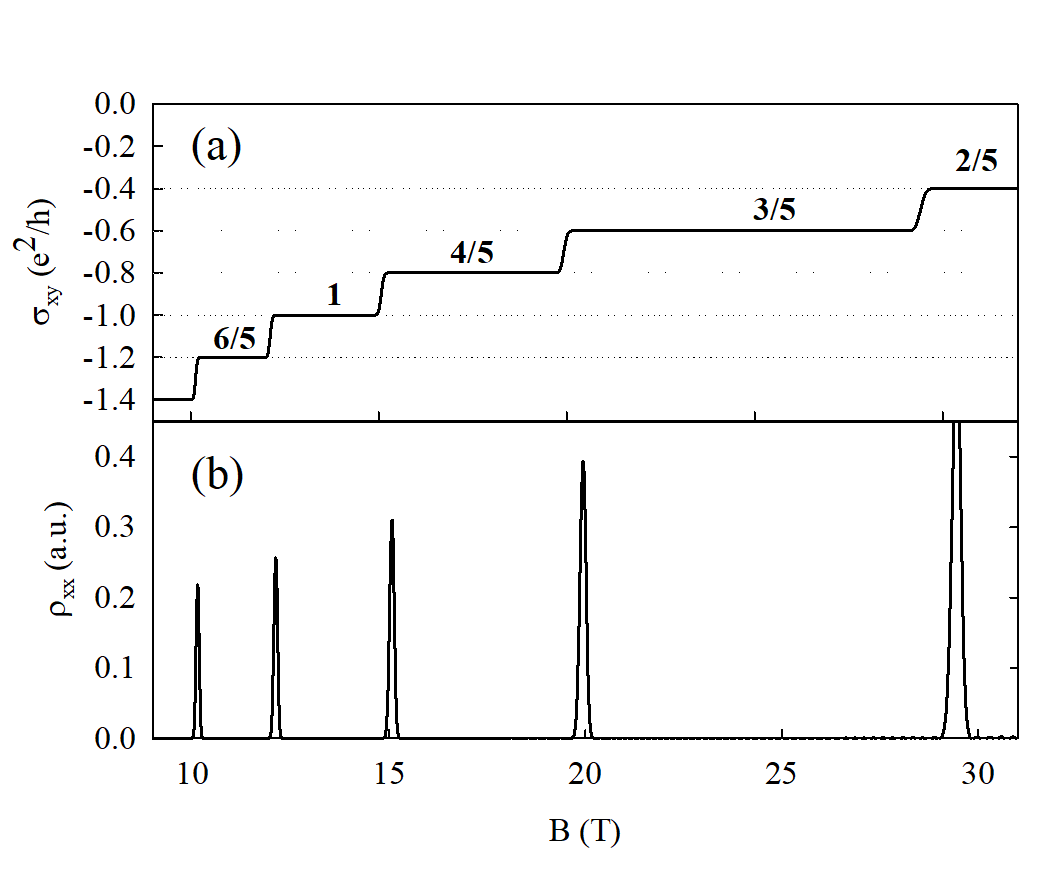}
\caption{Phenomenological fractional sublevel organization associated with higher-order odd-denominator states of the $1/5$ family. (a) Effective Hall magnetoconductivity exhibiting fractional plateau-like structures associated with the impurity-induced sublevels. (b) Corresponding longitudinal magnetoresistivity minima obtained from the fractional energy spectrum of Eq.~(\ref{eq:fractionalE}). The results illustrate how correlated orbital coherence can reorganize the effective Landau-level structure into experimentally relevant odd-denominator sequences.}
\label{fig:figure3}
\end{figure}

\bibliography{sn-bibliography}

\end{document}